\documentclass{article}
\def\usebibliography{}

\usepackage[english]{babel}
\usepackage[a4paper,top=2cm,bottom=2cm,left=3cm,right=3cm]{geometry}

\usepackage{amsmath,amssymb}     
\usepackage{graphicx}            
\usepackage[utf8]{inputenc}      
\usepackage{cleveref}            
\usepackage{caption}             
\usepackage{authblk}             
\usepackage{xcolor}              
\usepackage{setspace}            

\ifdefined\usebibliography
  \usepackage[authoryear, round]{natbib}
  \renewcommand{\cite}{\citep}  
\fi

\usepackage{titlesec}
\titleformat{\section}{\large\bfseries}{\thesection}{1em}{}
\titleformat{\subsection}{\mdseries\itshape\large}{\thesubsection}{1em}{}

\crefformat{figure}{#2Figure~#1#3}
\crefformat{table}{#2Table~#1#3}
\crefformat{section}{#2Section~#1#3}

\usepackage{wrapfig}
\usepackage{setspace}

\title{\fontsize{15.8}{20}{{\fontseries{sb}\selectfont Can an increase in productivity cause a decrease in production?\\
Insights from a model economy with AI automation\thanks{\copyright\;Casey O. Barkan, 2024. All rights reserved.}}
}}

\author{Casey O. Barkan
\thanks{Department of Physics \& Astronomy, University of California, Los Angeles. Email: barkanc@ucla.edu}
}

\date{Working Paper. November 22, 2024.}

\begin{document}
\newgeometry{top=0.9in, bottom=0.7in, left=1in, right=1in}
\begin{titlepage}
\begin{center}
    {\fontsize{14}{18}\selectfont \bfseries Can an increase in productivity cause a decrease in production?\\
Insights from a model economy with AI automation \par} 
    \vspace{0.5cm} 
    {\large  Executive Summary \par}
\end{center}
{\fontsize{11}{14}\selectfont 
        \setlength{\parskip}{1em} 
        \setlength{\parindent}{0pt} 

\textbf{Overview:} This work investigates an undesirable and counterintuitive economic scenario that could result from AI automation. It is shown that, for a certain economic structure, an increase in the productivity of automation technology will decrease total production (GDP). This provides a counterexample to the widespread assumption that productivity increases necessarily increase economic production. Although the economic model in this work does not capture the full structure of a real economy, the model illustrates how basic economic mechanisms can give rise to unexpected adverse outcomes. 
I argue that quantitative economic modeling should play a larger role in analyses of AI risks.
Such models also serve as testbeds to explore the effects of proposed government policies.

\textbf{Questioning a common assumption:} 
It is widely assumed that productivity increases due to AI will increase GDP. This assumption underlies redistributive policy proposals like Universal Basic Income (UBI), where it is assumed that redistribution will allow everyone to benefit from AI. While this assumption is supported by economic models with perfectly competitive markets, it is likely that a small number of AI companies will dominate the market for automation technology, undermining competition. Noncompetitive markets typically lead to adverse outcomes, and very little is known about how automation technology affects economies with imperfect competition.

\textbf{Could AI decrease economic production?}
This work shows that the answer is \textit{yes} for a certain economic structure. This is shown using a model economy with no competition in either product or labor markets, as illustrated schematically below. The firm chooses labor employment and production in order to maximize profit.
When AI productivity surpasses a critical
\begin{wrapfigure}[30]{r}{7cm}
\begin{center}
\includegraphics[width=6.5cm]{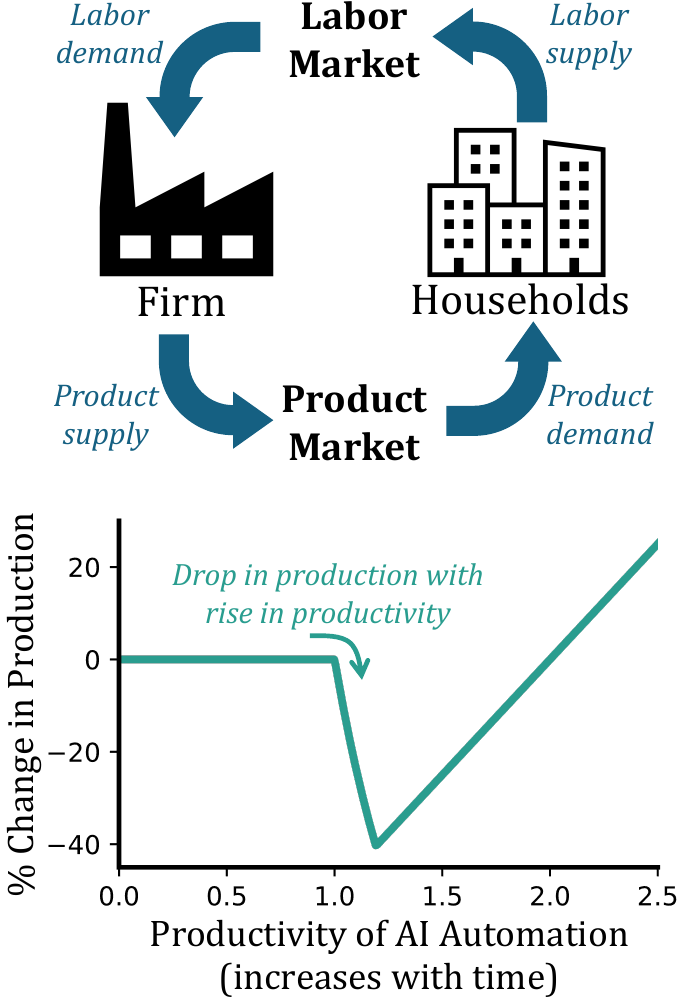}
\end{center}
\end{wrapfigure}
threshold, 
the firm replaces its labor force with AI, causing an increase in its profits but a decrease in total production, as shown in the figure.

\textbf{Models vs. Reality:}
Most of the literature on the macroeconomics of automation uses models with perfectly competitive markets, whereas my model assumes zero competition.
Reality is somewhere in between, and models of more realistic market structures are needed. There are other assumptions widely-used in macroeconomic models that I argue are not well suited for modeling AI impacts, including inelastic labor supply and representative agents.

\textbf{Outlook: AI Safety needs quantitative economic scenario modeling}.
Most economic research today is empirical, using historical data to understand the mechanisms that will shape the future. Yet, there is no data on the unprecedented changes that AI is likely to bring. This work illustrates how quantitative models can be used to explore scenarios for which no data exists, filling this gap.

}

\footnotetext[1]{Author: Casey O. Barkan}

\end{titlepage}
\restoregeometry
\clearpage

\maketitle
\onehalfspacing

\begin{abstract}
It is widely assumed that increases in economic productivity necessarily lead to economic growth. In this paper, it is shown that this is not always the case. An idealized model of an economy is presented in which a new technology allows capital to be utilized autonomously without labor input. This is motivated by the possibility that advances in artificial intelligence (AI) will give rise to AI agents that act autonomously in the economy. The economic model involves a single profit-maximizing firm which is a monopolist in the product market and a monopsonist in the labor market. The new automation technology causes the firm to replace labor with capital in such a way that its profit increases while total production decreases. The model is not intended to capture the structure of a real economy, but rather to illustrate how basic economic mechanisms can give rise to counterintuitive and undesirable outcomes.
\end{abstract}

\section{Introduction}
\label{sec:introduction}

Recent advances in artificial intelligence (AI) have inspired a growing public conversation on the extent to which AI will automate human labor, and on the economic impacts of automation. Nobel Laureate and AI pioneer Geoffrey Hinton has predicted that artificial general intelligence (AGI) will likely be developed by 2044\footnote{Hinton gave an interview in April 2024 at University College Dublin in which he said ``my guess is we'll get [AI] smarter than us, with a probability of about 0.5, in between 5 and 20 years." \cite{hinton2024interview}.}, and a recent survey of AI experts found that a majority expect AI to automate all human labor within the next 100 years \cite{grace2024thousands}. These advances have led to a flurry of research modeling the impacts of automation technology on economic growth, wages, and inequality \cite{aghion2017artificial, korinek2018artificial,acemoglu2018race, brynjolfsson2023macroeconomics,acemoglu2024simple, korinek2024scenarios, steigum2011robotics}. These studies include models that extend well-established macroeconomic models to incorporate automation technology. In doing so, they adopt many of the same assumptions used in these well-established models, notably, perfectly competitive markets and constant labor employment (equivalently, inelastic labor supply). Yet, AI will likely bring unprecedented changes to the economy, so assumptions that have been valid in the past may become invalid in the near future. In particular, AI may lead to a concentration of wealth in the hands of a few companies or individuals, which could undermine market competition.

In this paper, I use a simple model economy to demonstrate an undesirable and counterintuitive impact of automation that could occur if markets lack competition. Within the model, an increase in the \textit{productivity} of automation technology (this productivity is the output produced per unit of automation technology employed) causes a decrease in \textit{production} (total output, or GDP) when the economy has a certain structure. This result stands in contrast to the findings of prior models with competitive markets, which find the more intuitive result that increased productivity increases production. My model involves two markets, a product market and labor market, with a single firm which faces no competition in either market. The firm chooses the amount of human labor and capital\footnote{\textit{Capital} refers to the physical machinery used in production, which includes compute resources used by AI.} to employ in order to maximize profit. The firm has access to two technologies, an \textit{old} technology which requires both labor and capital input, and a new \textit{automation} technology which requires only capital input. I assume that the productivity of the automation technology starts from zero and increases, and I examine how production, wages, labor employment, and the firm's profit change as this productivity increases. When the productivity of the automation technology surpasses a threshold, the firm replaces its human labor force with automation technology. This replacement increases the firm's profit, but substantially decreases the firm's output. However, if the productivity of automation continues to rise beyond this threshold, total production will grow and eventually surpass its original level.

The model in this work is too simple to quantitatively describe a real economy, and one must ask, \textit{what is the value of such a simple model?} I argue that the value is twofold. First, highly simplified models play a central role in economics, forming the foundations upon which more sophisticated models are built. For example, the Ramsey model of economic growth \cite{ramsey1928mathematical}, which involves only two markets, a representative firm, and a representative agent, serves as the foundation for a large portion of modern macroeconomic models \cite{ljungqvist2018recursive}. Typically, the key insights from the simplified models generalize to the more sophisticated models. With regard to the model presented here, extensions of this model to more realistic economic structures with imperfectly competitive markets may clarify whether the phenomenon of decreasing production with increasing productivity could occur in the real economy. Second, this work shows a \textit{counterexample} to a prevailing assumption, namely, that productivity increases necessarily lead to production increases. Counterexamples are crucial for motivating a re-examination of assumptions, for refining intuitions and conceptual understandings, and ultimately for guiding improved models and theories.


In AI safety and governance research, threat modeling and scenario modeling are important tools for exploring the range of harmful and beneficial impacts that AI may bring. The trajectory of AI's capabilities, dangers, and societal influence are highly uncertain, so models that explore a wide range of scenarios are needed. I would argue there is a similar need for economic scenario modeling. AI's future impacts on the economy are highly uncertain, but the impacts are likely to be profound. Hence, there is a need for economic models that deviate from standard modeling paradigms and that explore less common economic structures and counterintuitive mechanisms. The model in this work is meant to be a first step toward such economic scenario modeling.



\section{Model}

\begin{figure}
 \centering
 \makebox[\textwidth][c]{\includegraphics[width=\textwidth]{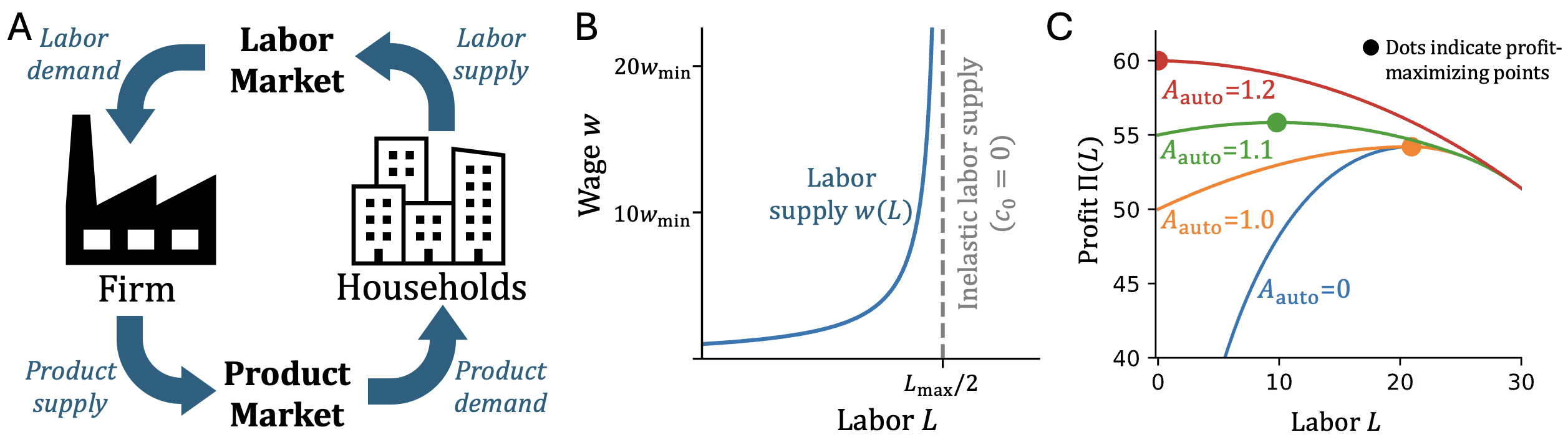}}%
 \caption{\textbf{Model structure, labor supply, and profit maximization} (A) Schematic of the structure of the economy. (B) Labor supply with $c_0>0$ (blue), and an inelastic labor supply ($c_0=0$) for comparison (dashed grey). (C) Firm's profit function $\Pi(L)$ for increasing values of the productivity $A_\text{auto}$. As $A_\text{auto}$ increases beyond a value of 1, the profit-maximizing labor $L^*$ (indicated by dots) jumps from $L^*\approx 20$ down to 0. Model parameters used to generate plots are listed in section \ref{computing}.}
 \label{model}
\end{figure}

Consider a model economy with two markets: a product market and a labor market, as illustrated schematically in Fig.~\ref{model}A. Labor is assumed to be undifferentiated and all laborers receive the same wage. There is a single firm which is a monopolist in the product market and a monopsonist in the labor market\footnote{This means that they are the only firm operating in either market, facing no competitors.}. The firm's ability to produce is described by a production function (defined below), which specifies the amount of product that the firm can produce as a function of labor and capital input. The production function depends upon the scientific development of AI automation technology; specifically, a parameter $A_\text{auto}$ specifies the productivity of automation technology, which is assumed to increase with time as scientific progress in AI is made\footnote{To be concrete, $A_\text{auto}$ captures advances in AI architectures as well as advances in performance (i.e. improved model weights) for given architectures.}. The households that supply labor choose their consumption and labor supply by maximizing a utility function, and all households are assumed to have the same utility function. Lastly, it is assumed that the households that supply labor do not own capital or equity in the firm, so that wages are their only source of income. The general equilibrium is computed as a function of $A_\text{auto}$ to determine how the economy's production (GDP), labor employment, and wages, depend upon the productivity of automation technology.

The total capital stock in the economy is assumed to be fixed at a constant value $\bar K$. This is a valid approximation when technological change occurs rapidly relative to the rate of capital accumulation. Whether this approximation is valid for AI automation technology remains to be seen, though the model can be extended to include capital accumulation. In fact, dynamic growth models of economies with competitive markets and automation technology have been developed previously, and they predict that automation produces unbounded endogenous growth \cite{steigum2011robotics, aghion2017artificial, prettner2020automation}. It is an interesting open question how this prediction would change for an economy with noncompetitive markets.

\subsection{The firm's production decision}

The firm chooses its production $f$, labor employment $L$, capital employment $K$, and wage $w$ so as to maximize profit. The firm has access to two technologies for production, the old technology and the automation technology, and the firm chooses how to allocate capital between the two technologies to maximize profit. Each technology has a production function of the Cobb-Douglas form \cite{barro2004economic}. The old technology has production function $f_\text{old}(K,L)=A_\text{old}
K^\alpha L^{1-\alpha}$ with $0<\alpha<1$. The automation technology has production function 
$f_\text{auto}(K)=A_\text{auto}K$, which is a Cobb-Douglas function with exponent equal to 1. These production functions have the following key properties: with the old technology, capital requires labor input to be productively utilized, whereas, with the automation technology, capital is productive without labor input. $A_\text{auto}$ represents the productivity of the automation technology, and I assume that $A_\text{auto}=0$ initially, then study how the economy changes as $A_\text{auto}$ increases.

The firm's total production function, $f(K,L)$, is determined by allocating capital between the two technologies so as to maximize output. Mathematically,
\begin{equation}\label{f_max}
    f(K,L) = \max_{K_\text{old}\in[0,K]} \; \big( f_\text{old}(K_\text{old},L) + f_\text{auto}(K-K_\text{old}) \big)
\end{equation}
The $K_\text{old}$ that solves this maximization problem can be found by differentiating the expression being maximized with respect to $K_\text{old}$, setting the result to zero, and solving for $K_\text{old}$. This yields the optimal capital allocation $K^*_\text{old}(K,L)$, given by
\begin{equation}\label{K_old*}
    K^*_\text{old}(K,L) = \min\left\{L\left(\frac{\alpha \,A_\text{old}}{A_\text{auto}}\right)^\frac{1}{(1-\alpha)}, K\right\}
\end{equation}
The minimum operation in Eq.~\ref{K_old*} ensures that the constraint $K_\text{old}\in[0,K]$ is satisfied. The production function can now be rewritten as
\begin{equation}\label{f}
    f(K,L) = A_\text{old}K_\text{old}^*(K,L)^\alpha L^{1-\alpha} + A_\text{auto}\left(K-K_\text{old}^*(K,L) \right)
\end{equation}

To express the firm's profit as a function of $K$ and $L$, we use the price of product as the numeraire\footnote{The \textit{numeraire} in an economic model is the unit according to which all prices are measured, and it is standard to use the price of the representative product as the numeraire \cite{barro2004economic}. This means that, for example, if wage $w=2$, then two units of product can be purchased with the wage earnings from one unit of labor.} (meaning units are set so the price of product equals 1), so profit $\Pi$ is given by
\begin{equation}\label{Pi}
    \Pi(K,L) = f(K,L) - w(L)L - r(K)K
\end{equation}
where $w(L)$ is the wage that the firm must offer to procure labor quantity $L$ in the labor market ($w(L)$ is the \textit{labor supply curve}), and $r(K)$ is the rental rate of capital. The firm chooses, $f$, $L$, and $K$ in order to maximize $\Pi(K,L)$.

Lastly, I assume that it is profit maximizing for the firm to utilize the entire capital stock\footnote{This is a very minor assumption for the following reason: The only case in which this assumption would fail to hold is if the operating cost of capital is so high that some capital is left unused. We would not expect this to occur, because such capital would not have been manufactured if it were too expensive to be profitably used.}. 
This is valid if the maximum rental rate $\bar r \equiv r(\bar K)$ is lower than the marginal product of capital at the profit-maximizing point. This simplifies the profit function to a function of only $L$,
\begin{equation}\label{Pi_of_L}
    \Pi(L) = f(\bar K,L) - w(L)L - \bar r \bar K
\end{equation}
and the firm chooses $L$ to solve $\max_L \Pi(L)$. To solve this profit maximization problem, we need to know the labor supply $w(L)$. In the section below, the households' utility maximization problem is solved to derive the supply curve,
\begin{equation}\label{w_of_L}
    w(L) = \frac{w_\text{min}}{1-L/(\gamma L_\text{max})}
\end{equation}
where $\gamma$ is a parameter in the utility function, as described below. This labor supply curve is shown in Fig.~\ref{model}B.

The problem $\max_L \Pi(L)$ can now be solved to obtain the equilibrium labor $L^*$, production $f^*=F(\bar K,L^*)$, and profit $\Pi^*=\Pi(L^*)$. In equilibrium, the firm allocates an amount of capital $K_\text{old}^*(\bar K,L^*)$ to the old technology, and the remainder of the capital stock is allocated to the automation technology.

Fig.~\ref{model}C shows the profit function $\Pi(L)$ and the profit maximizing point (indicated by a dot) for four values of $A_\text{Auto}$. As shown, the profit maximizing labor $L^*$ remains fixed until $A_\text{Auto}>1$, then it quickly drops and reaches zero when $A_\text{Auto}\gtrsim1.2$, indicating that the labor force is replaced by AI automation.

\subsection{Households' labor and consumption decision}

Nearly all prior studies on the impacts of automation assume inelastic labor supply, which implies constant labor employment\footnote{I am aware of only one exception to this: \cite{prettner2017lost}}. Yet, displacement of labor is one of the primary concerns regarding the economic impacts of AI. In this model, I do not assume inelastic labor supply; rather, I model labor supply by assuming that households select their labor and consumption quantities to maximize a utility function $U(c,\ell)$, where $c\equiv wL$ is consumption (equal to labor income) and $\ell\equiv L_\text{max}-L$ is leisure, where $L$ is the labor supplied by households and $L_\text{max}$ is the maximum amount of labor that households can provide. Assume that $U(c,\ell)$ takes the form
\begin{equation}
    U(c,\ell) = (c+c_0)^{\gamma}\ell^{1-\gamma}
\end{equation}
where $\gamma$ is a parameter that specifies households' preference for consumption over leisure, and $0<\gamma<1$. The parameter $c_0$ captures an essential aspect of the labor supply, namely, that there exists a minimal wage $w_\text{min}$ below which households do not supply labor. $w_\text{min}$ is not a legally-imposed minimum wage, rather, it reflects the fact that if wages are too low, households will choose not offer labor or will be incapable of offering labor due to the wage not providing subsistence-level consumption. If $c_0=0$ then this utility function produces an inelastic labor supply curve. However, if $c_0>0$ or $c_0<0$, then the resulting labor supply is not perfectly inelastic, and there will be a minimal wage $w_\text{min}$, as derived below. There are multiple ways to interpret $c_0$. It could be merely an intrinsic parameter that characterizes households preferences. Or, if $c_0>0$, it could correspond to an exogenous source of consumption, perhaps due to a welfare policy like universal basic income. $c_0<0$ could describe a scenario where households require a minimal consumption level to subsist, so marginal utility approaches infinity in the limit that $c$ drops to $c_0$, and households cease to provide labor for $c\leq c_0$ because starvation occurs.

We will solve the households' utility maximization problem for an arbitrary $c_0\neq 0$. The maximization problem is
\begin{equation}\label{utility_prob}
\begin{split}
    \max_{c,\ell} \; U(c,\ell) \quad &\text{subject to} \quad c = w(L_\text{max}-\ell) \\
    &\text{and} \quad c \geq 0, \quad c>-c_0, \quad \ell>0
\end{split}
\end{equation}
Solving this using a Lagrange multiplier and setting $\ell=L_\text{max}-L$ yields the labor supply $w(L)$, which is the wage that induces households to offer $L$ units of labor. The result is
\begin{equation}\label{full_labor}
    w(L) = \frac{(1-\gamma)c_0}{\gamma L_\text{max}-L}
\end{equation}
This can be written in a more intuitive form in terms of $w_\text{min}$, as follows:
\begin{itemize}
    \item For $c_0>0$, $w_\text{min}=\frac{1-\gamma}{\gamma}\frac{c_0}{L_\text{max}}$ and 
    \begin{equation}
        w(L) = \frac{w_\text{min}}{1-L/(\gamma L_\text{max})}
    \end{equation}
    This labor supply curve is shown in Fig.~\ref{model}B.
    \item For $c_0<0$, $w_\text{min}=-c_0/L_\text{max}$ and
    \begin{equation}
        w(L) = \frac{1-\gamma}{\gamma}\frac{w_\text{min}}{1-L/(\gamma L_\text{max})}
    \end{equation}
    This labor supply curve is \textit{downward sloping} for $w>w_\text{min}$. For $w\leq w_\text{min}$, the household cannot meet the subsistence level of consumption $c_0$; in other words, starvation occurs, so labor supplied drops to 0. Despite the grim interpretation of this scenario, the model's predictions of production, labor employment, profit, and capital allocation are nearly identical to the predictions for $c_0>0$.
\end{itemize}

One might wonder why we have not computed a demand curve for the product. In fact, the product demand is implicitly defined by Eq.~\ref{full_labor} and by our choice to use the product price as the numeraire. We could solve the problem differently, by setting the wage as the numeraire, in which case solving Eq.~\ref{utility_prob} would yield a product demand curve instead of a labor supply curve.

\subsection{Computing the equilibrium}\label{computing}

Due to the monopolist-monopsonist structure of the economy, the equilibrium is determined by the firm's profit maximization problem, i.e. by maximizing Eq.~\ref{Pi_of_L}. The labor market clearing condition is $w=w(L)$ and product market clearing condition is $f=wL+rK-\Pi$, and both of these conditions are satisfied by maximizing Eq.~\ref{Pi_of_L}. To generate Figures 1 and 2, I used the following parameters: $\alpha=0.5$, $\gamma=0.5$, $w_\text{min}=2$, $\bar K=50$, and $L_\text{max}=500$. $A_\text{old}$ is set so that the marginal product of capital equals 1 when $A_\text{auto}=0$, which yields $A_\text{old}\approx 3.01$. The maximization of Eq.~\ref{Pi_of_L} is done numerically and code is provided\footnote{Python code is available at https://github.com/cbarkan1/socioeconomic-impacts-of-AI}.

\section{Results}

\begin{figure}
 \centering
 \makebox[\textwidth][c]{\includegraphics[width=\textwidth]{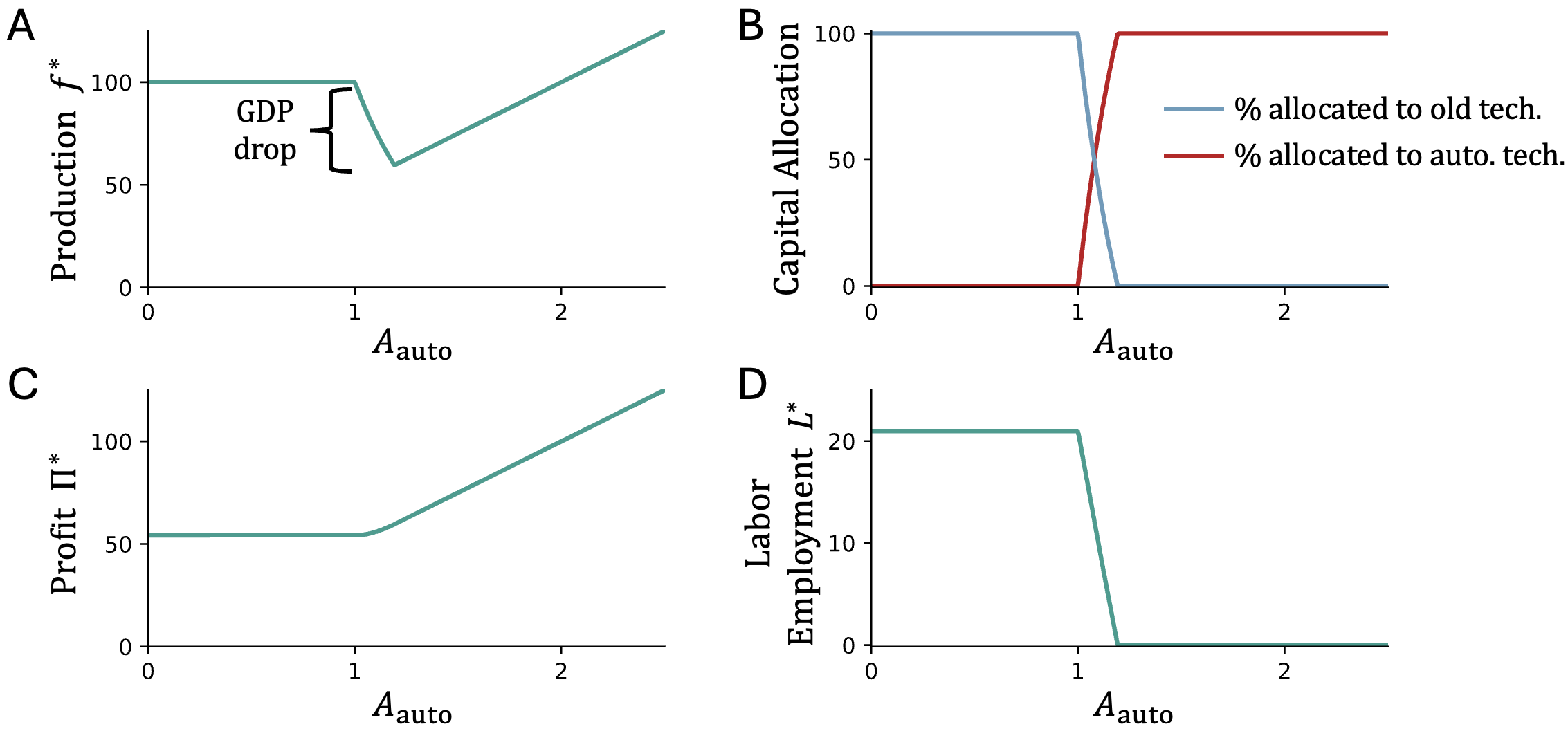}}%
 \caption{\textbf{Equilibrium quantities versus the productivity of automation technology $A_\text{auto}$.} (A) Production ($f^*$) vs. $A_\text{auto}$. (B) Percent capital allocation vs. $A_\text{auto}$. Blue and red curves show, respectively, percent of capital allocated to the old and automation technologies. (C) Firm's profit ($\Pi^*$) vs. $A_\text{auto}$. (D) Labor employment ($L^*$) vs. $A_\text{auto}$.}
 \label{equilibrium}
\end{figure}

Suppose that the productivity of automation technology, $A_\text{auto}$, equals 0 initially, and $A_\text{auto}$ increases as automation technology develops. When $A_\text{auto}$ is low, it is most profitable for the firm to only utilize the old technology. But, once $A_\text{auto}$ surpasses the marginal productivity of capital of the old technology, the firm begins to reallocate capital from the old technology to the new automation technology\footnote{The initial marginal productivity of capital of the old technology is $MP_{k0}\equiv\partial_Kf_\text{old}(\bar K,L^*)|_{A_\text{auto}=0}$, and model parameters are chosen so that $MP_{k0}=1$. Hence, once $A_\text{auto}$ surpasses 1, the transition begins, as shown in Fig.~\ref{equilibrium}.}. This transition is illustrated in Fig.~\ref{equilibrium}, where the transition occurs at $A_\text{auto}=1$.

Fig.~\ref{equilibrium} panels A, B, C, and D show, respectively, how production, capital allocation, profit, and labor employment, change as $A_\text{auto}$ increases. As capital is reallocated from the old technology to the new, a reduction in labor employment causes a decrease in production $f$ but a \textit{larger} decrease in labor costs $wL$. Because profit $\Pi=f-wL-rK$, it is profitable for the firm to decrease $f$ if, in doing so, it decreases $wL$ by a greater amount. This is exactly what the firm does: it decreases its labor employment to zero, decreasing production while drastically decreasing costs, leading to an increase in profit despite less total product being produced in the economy.

The abruptness of this transition is concerning from a policy perspective. As AI automation technology develops, the model predicts no change in the economy until a massive and abrupt transition occurs. In other words, AI technology can become quite advanced with no societal impact, until the technology reaches a critical level, at which point a profound restructuring of the economy occurs.

After the transition to the automation technology occurs, production begins to rise with $A_\text{auto}$, eventually surpassing its pre-transition level. If $A_\text{auto}$ continues to grow to be arbitrarily large, production $f$ will become arbitrarily large as well (in fact $f$ and $A_\text{auto}$ are linearly related after the transition). Although all labor has been displaced in this post-transition scenario, a redistributive policy like universal basic income would let everyone benefit from the transition once $A_\text{auto}$ becomes sufficiently large. However, even if the drop in production is transient and a redistributive policy is enacted, the transient drop could have severe and lasting consequences (for instance, people may starve when labor is displaced). Moreover, if a large drop in production were to occur, it would likely induce a financial crisis which could hinder the technological developments that lead to continued growth of $A_\text{auto}$. In this case, the economy may become stuck in the regime of reduced production.

With the model parameters used in Fig.~\ref{equilibrium}, an approximately  40\% drop in production occurs. This is substantially larger than the drop in GDP during the Great Depression, which, in the United States, was 30\%. However, the magnitude of the drop depends on the model parameters, and there are certain parameter values for which no drop in production occurs. Ideally, one could estimate model parameters from data, but the model in this work is too simplistic to be fit to real-world data.

\section{Comparison with other models}

The model in this work has several differences from recently published macroeconomic models that incorporate automation. Some of these differences give this model a unique perspective, but others are limitations that can be improved upon in future work.
\begin{itemize}
    \item \textbf{Competition.} The lack of competitive markets in this model is the key feature that makes it unique from other macroeconomic models of automation. There is only one other published model that studies automation in imperfectly competitive markets \cite{acemoglu2024automation}. Their model involves labor markets where certain workers are paid above opportunity cost. The implications of their model are interesting but completely distinct from the implications of the model in this work.
    
    \item \textbf{Investment and Capital Accumulation.} The accumulation of capital via investment is a central feature of models of economic growth. Capital accumulation is omitted from the model in this work for simplicity, but the model can be extended to include capital accumulation in a straightforward way. When capital accumulation of automation technology occurs, economies can achieve unbounded growth \cite{steigum2011robotics,prettner2017lost, aghion2017artificial}.
    
    \item \textbf{Labor Supply.} The model in this work involves an elastic labor supply curve, and the labor supply captures the fact that workers will not supply labor when the wage is below a minimal value. This is essential for modeling labor displacement. This contrasts from nearly all other macroeconomic models of automation, where constant labor employment (inelastic labor supply) is assumed. I am aware of one other work that incorporates elastic labor supply to model labor displacement due to automation \cite{prettner2017lost}.
    
    \item \textbf{Approach to modeling automation technology. } There are multiple ways in which automation can be incorporated into macroeconomic models. For the model in this work, I assume that there is an existing capital stock which can be repurposed for automation, and the productivity of capital utilized for automation is described by a parameter $A_\text{auto}$ which increases with time. To explicitly relate this to real AI, capital would refer to compute resources and $A_\text{auto}$ would include trained model weights which improve as better AI models are trained. Of course, compute resources are not the only form of capital in the economy, so this approach is an oversimplification. A different approach, taken in \cite{prettner2020automation}, is to assume automation capital is an entirely distinct form of capital, and existing capital cannot be repurposed for automation. In this approach, the automation capital stock begins at zero and accumulates over time via savings. This approach is also an oversimplification, because in the real world, new AI models do not require new compute resources, rather new AI models are run on existing hardware. An improved model could take the best of both of these approaches, by treating compute resources as its own category of capital which must accumulate (as in the model in \cite{prettner2017lost}) and which can be reallocated between AI models with time varying productivities (as in my model).
    
    There is a more fine-grained approach to modeling automation used in \cite{acemoglu2018race, acemoglu2024simple, korinek2024scenarios}. In these works, the production function is expressed in terms of a continuum of tasks, and tasks are completed either by human labor or by capital. This approach allows the model to capture the fact that certain tasks are easier to automate than others, yet it also does not differentiate between different forms of capital.
\end{itemize}

\section{Discussion}

An essential question is, who benefits and who is harmed by the transition to automation in this model? The owners of the firm and the owners of capital benefit, as their profits increase through the transition to automation despite declining production. Laborers, whose employment drops to zero during the transition, are harmed. In the absence of redistributive policy, the displaced laborers must either subsist through non-market activities or starve. Redistributive policies like universal basic income (UBI) are widely proposed as a means to support displaced laborers. Yet, if AI automation leads to a decrease in GDP, redistributive policies cannot remedy the fact that there is less total wealth to be distributed.

Models like the one in this work provide a setting in which to explore the effects of government policies. As an example, a redistributive policy coupled with a tax policy could be incorporated into the model in the following way. Consider a policy in which the firm is obligated to make direct welfare payments to unemployed workers; this is not a fixed tax rate, but a dynamic tax that varies as unemployment varies. With such a policy, the firm is disincentivized from replacing labor with automation. Incorporating such a policy into the model is left for future work, but I hypothesize that the firm will not replace its human labor force with automation until the productivity of automation is sufficiently high that no drop in production occurs when the transition is made. In general, policies can be incorporated into the mathematical framework of economic models, so that the models can make predictions about the policies' effects.

Extensions of the model to more realistic contexts are needed to make predictions about the real economy. Some features to include in subsequent models are:
\begin{itemize}
    \item Imperfectly competitive markets, which describe a middle-ground between perfect competition and zero competition.
    \item Differentiated capital (i.e. different categories of capital). In particular, compute resources should be treated as distinct from other forms of capital. Advances in AI technology would correspond to an increase in the productivity of compute, but may not affect the productivities of other forms of capital.
    \item Capital accumulation and dynamic growth, describing the accumulation of compute resources through time.
    \item Heterogeneity among agents. In real economies, people differ in their capital ownership and labor skillset, and these differences can be described with heterogeneous agent models \cite{guvenen2011macroeconomics}. Heterogeneous agent models also allow for modeling of financial markets in which different agents have different investment behaviors. In a world with diminishing labor income, returns on financial investments may become a larger source of income for a growing portion of the population, so such models may be especially relevant. Heterogeneous agent models can also describe the dynamics of the distribution of wealth in the economy, which will likely become increasingly concentrated as AI develops.
\end{itemize}

Stepping back to look at economic research broadly, most research today uses empirical data to investigate economic mechanisms and to forecast the future. This empirical data reflects the structure of our current society, yet its relevance for a future reshaped by AI is limited. As an alternative, theoretical models provide a way to explore scenarios for which no empirical data exists. For this reason, economic models incorporating AI automation are essential.

\bibliographystyle{plainnat}  
\bibliography{main}

\end{document}